\documentclass[journal=jacsat,manuscript=article]{achemso}

\usepackage{chemformula} 
\usepackage[T1]{fontenc} 
\usepackage{ulem}

\author{Sara Freund}
\affiliation{University of  Basel, Department of Physics, Klingelbergstrasse 82, CH-4056 Basel, Switzerland}
\author{R\'emy Pawlak}
\affiliation{University of  Basel, Department of Physics, Klingelbergstrasse 82, CH-4056 Basel, Switzerland}
\email{remy.pawlak@unibas.ch}
\author{Lucas Moser}
\affiliation{University of  Basel, Department of Physics, Klingelbergstrasse 82, CH-4056 Basel, Switzerland}
\author{Antoine Hinaut}
\affiliation{University of  Basel, Department of Physics, Klingelbergstrasse 82, CH-4056 Basel, Switzerland}
\author{Roland Steiner}
\affiliation{University of  Basel, Department of Physics, Klingelbergstrasse 82, CH-4056 Basel, Switzerland}
\author{Nathalie Marinakis}
\affiliation{University of  Basel, Department of Chemistry, Mattenstrasse 24a, BPR 1096, CH-4058 Basel, Switzerland}
\author{Edwin~C. Constable}
\affiliation{University of  Basel, Department of Chemistry, Mattenstrasse 24a, BPR 1096, CH-4058 Basel, Switzerland}
\author{Ernst Meyer}
\affiliation{University of  Basel, Department of Physics, Klingelbergstrasse 82, CH-4056 Basel, Switzerland}
\author{Catherine~E. Housecroft}
\affiliation{University of  Basel, Department of Chemistry, Mattenstrasse 24a, BPR 1096, CH-4058 Basel, Switzerland}
\author{Thilo Glatzel}
\affiliation{University of  Basel, Department of Physics, Klingelbergstrasse 82, CH-4056 Basel, Switzerland}

\title[An \textsf{achemso} demo]
{Transoid-to-Cisoid Conformation Changes of Single Molecules on Surfaces Triggered by Metal Coordination}

\begin{document}

\begin{abstract}
Conformational isomers are stereoisomers that can interconvert over low potential barriers by rotation around a single bond. However, such bond rotation is hampered by geometrical constraints when molecules are adsorbed on surfaces. Here we show that the adsorption of 4,4$'$-bis(4-carboxyphenyl)-6,6$'$-dimethyl-2,2$'$-bipyridine molecules on surfaces leads to the appearance of pro-chiral single-molecules on NiO(001) and to enantiopure supramolecular domains on Au(111) surfaces containing the $transoid$ molecule conformation. Upon additional Fe adatom deposition, molecules undergo a controlled interconversion from a $transoid$ to $cisoid$ conformation as a result of coordination of the Fe atoms to the 2,2$'$-bipyridine moieties. As confirmed by atomic force microscopy images and X-ray photoelectron spectroscopy measurements, the resulting molecular structures become irreversibly achiral.
\end{abstract}

\section{Introduction}
\begin{figure}[t]
 \centering
 \includegraphics[width=14cm]{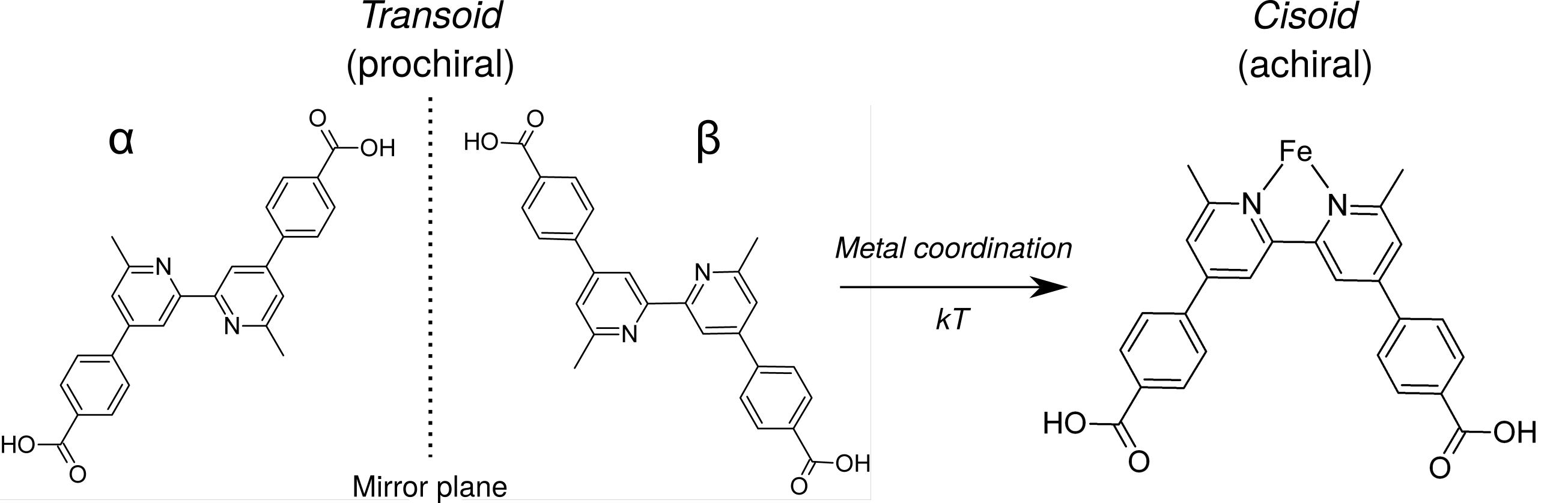}
 \caption{ $Transoid$ and $Cisoid$ Conformations of The DCPDMbpy And Fe-DCPDMbpy Molecules. Upon adsorption, the DCPDMbpy adopts two prochiral conformations $\alpha$ and $\beta$. Upon coordination with Fe, the molecule can undergo a conformational change to the $cisoid$-conformation.}
 \label{Fig1}
\end{figure}

\indent
An enantiomer is "one of a pair of molecular entities which are mirror images of each other and non-superposable"~\cite{IUPAC1}. Atropisomerism is a particular class of axial enantiomerism which results from hindered rotation about a single bond. In such compounds, enantiomer interconversions are mediated only by bond rotations between isomers (in contrast to interconvertions that involve covalent bond breaking). Thus, the stability of "long-lived" atropisomers in three-dimensions usually requires steric hindrance in order to constrain internal bond rotations using peripheral chemical substitutions .  
To impose chirality, another approach consists in the confinement of molecules onto a crystalline surface. Over the last couple of decades, this strategy has enabled the formation of enantiopure self-assemblies~\cite{Rav09, Ern12} or chiral molecular compounds from on-surface chemical reactions~\cite{Ste16,Wac16}.  Accessing chiral molecular surfaces further allows a vast range of novel properties to emerge including the amplification of non-linear optical properties~\cite{Ver98,Ros08} and the asymmetric scattering of spin-polarized electrons~\cite{Ray99}. Moreover, the control of chiral-achiral transitions in surface-stabilized molecular networks could also help designing chirality sensors, molecular switches and motors~\cite{Kot05,Bro06,McC01}. If a good alternative for stabilizing relies in their geometrical frustration on a surface~\cite{Oht99, Hin18}, a step further would be to control the bond rotation and thus the molecule conformation.   \\ 
\indent
In this work, we investigate the adsorption of achiral 4,4$'$-di(4-carboxyphenyl)-6,6$'$-dimethyl-2,2$'$-bipyridine molecules (DCPDMbpy) by means of atomic force microscopy (AFM), scanning tunneling microscopy (STM) and X-ray photoelectron spectroscopy (XPS) on NiO(001) and Au(111). Our work is motivated by our recent hierarchical assembly strategy "surfaces-as-ligands surfaces-as-complexes"  (SALSAC) approach~\cite{Hou15} focusing on designing novel molecular compounds having {\it (i)} anchoring groups such as carboxylic or phosphonic acids that enable a strong anchoring of the molecule to surfaces and {\it (ii)} metal-binding moieties such as 2,2$'$-bipyridine ($bpy$) to facilitate the assembly of surface-bound metal coordination compounds either through sequential addition of metal ions and an ancillary ligand, or through a ligand-exchange reaction between the anchoring ligand and a homoleptic metal complex~\cite{Hou15, Sch15,Boz11,Mal17}.\\
\indent
 In a previous work, we investigated the first step of the DCPDMbpy assembly process using AFM operated in ultrahigh vacuum (UHV) conditions.~\cite{Fre17} 
We showed that the DCPDMbpy molecule systematically adopts two prochiral $transoid$-conformations ($\alpha$ and $\beta$, Figure~\ref{Fig1}a) when adsorbed onto an atomically clean NiO(001) crystal surface. In contrast, the $cisoid$-conformation was not observed even upon annealing up to 420~K close to the desorption temperature of the molecule. This can be explained by the high energy barrier needed to be overcome in order to induce a bond rotation about the interannular C--C bond as well as the energy to partially desorb the molecule from the surface to allow the rotation.\\
\indent
Theoretical studies of the conformational change from $transoid$ to $cisoid$ for a $bpy$ in the gas phase have estimated an energy of about 320~meV~(31~kJ$\cdot$mol$^{-1}$)~\cite{Goe00, How96}. The high rotational barrier was found to arise from the electrostatic repulsion between the lone pairs of the bipyridine units~\cite{Gut18}. This repulsive interaction also leads to the transoid conformation in favor of the cisoid one in gas phase as well as during its adsorption on surfaces. Note that this barrier is much higher than the available thermal energy at room temperature or even upon surface annealing ( i.e. at 1000K, the available energy is about 80~meV~(8~kJ$\cdot$mol$^{-1}$) which implies that the molecule desorb before changing its conformation to the $cisoid$ one). However, the activation energy to promote the conformational change can be overcome when forming a complex between a metal atom and a $bpy$ unit, theoretically delivering about 4.66~eV~(450~kJ$\cdot$mol$^{-1}$)~\cite{Ama17, Rod00}. Metal coordination might thus enable the emergence of the $cisoid$ form on surfaces.\\
\indent
Here, we investigated DCPDMbpy molecules in the presence of metallic adatoms on both NiO(001) and Au(111) surfaces. The achiral $cisoid$ geometry is observed after adsorption of prochiral DCPDMbpy on a NiO(001) surface previously partially covered with Fe atoms, demonstrating the Fe-DCPDMbpy complex formation through coordination between Fe atoms and $bpy$ moieties.
Furthermore, upon Fe adatom deposition on prochiral DCPDMbpy assemblies formed on Au(111), the molecules undergo an interconversion from $transoid$ to $cisoid$ on the surface, which is triggered by the same metal-coordination mechanism as shown by AFM images and XPS measurements.    

\section{Results and discussion}

\subsection{The Conformations of The DCPDMbpy And Fe-DCPDMbpy Molecules on NiO(001) at Room Temperature.} 
 
\begin{figure}[t]
 \centering
 \includegraphics[width=16cm]{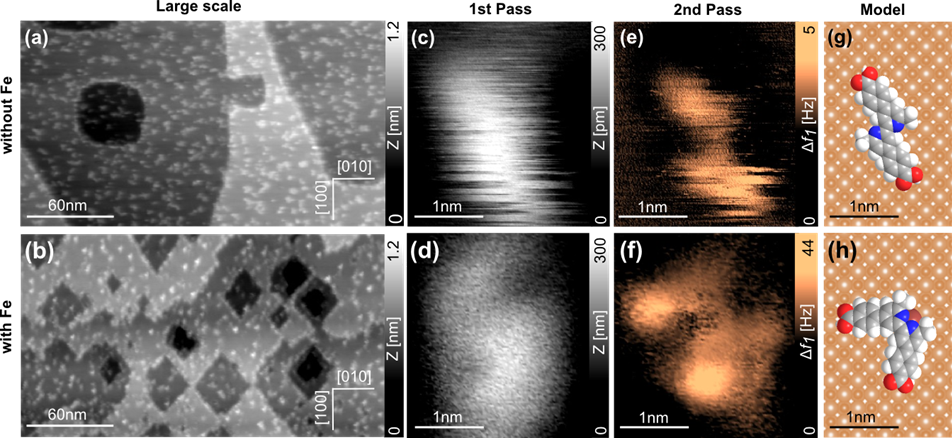}
 \caption{$Transoid$- And $Cisoid$-Conformations of DCPDMbpy Molecules Adsorbed on NiO(001). Large-scale AFM topographic images of NiO(001): (a) after deposition of DCPDMbpy molecules and (b) after deposition of Fe and DCPDMbpy molecules (scan parameters: $A_{1} = 4$~nm, $\Delta f_{1} = -3$~Hz and $\Delta f_{1} = -10$~Hz, respectively). (c) and (d) AFM topographic images of flat-lying single-molecules adsorbed before and after Fe deposition, respectively. The images were acquired using the first scanning pass (scan parameters: $A_{1} = 4$~nm, $\Delta f_{1} = -2.5$~Hz and $\Delta f_{1} = 29$~Hz, respectively). (e) and (f) Corresponding $\Delta f$ images acquired with the second scanning pass, i.e. with open feedback, using offsets of $\sim$ -350 pm and -280 pm.  The molecules are in $transoid$- and $cisoid$- conformation, respectively. (g) and (h) structural models of both DCPDMbpy geometries on NiO(001).}
  \label{Fig2}
 \end{figure}

To demonstrate the control of such chiral-achiral transitions on surfaces, we first investigated  DCPDMbpy molecules on a NiO(001) substrate. Figure~\ref{Fig2}a shows a representative AFM topographic image of the NiO(001) surface after deposition of 0.2 monolayer of DCPDMbpy at room temperature (RT). In the following, we define a monolayer (ML) as one layer of molecules fully covering the surface, 0.2~ML corresponds to a surface coverage of 20$\%$. Large terraces are separated by mono-atomic steps and are covered with single molecules as well as molecular aggregates. The step edges are saturated on both upper and lower sides indicating that these are preferential adsorption sites. The relatively short distance between the molecules (3.9 $\pm$ 0.7 nm in average) suggests a relatively low diffusion of the molecules at RT on the NiO(001) surface and, therefore, a rather strong binding to the substrate. Upon surface annealing, the diffusion remains low as discussed previously~\cite{Fre17}, where molecular diffusion as a function of the substrate annealing temperature was studied.

In order to trigger the emergence of metal complexes, 0.1~ML of Fe atoms were deposited at RT on the bare surface of NiO. DCPDMbpy molecules were then subsequently adsorbed onto this surface. To favour the coordination complex formation, the sample was annealed to 420~K after sublimation of molecules. Figure~\ref{Fig2}b shows an AFM topographic image of this surface.
To unambiguously confirm the DCPDMbpy-Fe complex formation on NiO(001), we focused on imaging of the molecule conformations at RT using a silicon cantilever (see Methods), employing the multipass technique~\cite{Mor15}  which has proven to deliver submolecular resolution at RT~\cite{Iwa15,Fre17}. The method consists of recording a first line scan with a closed feedback loop at a relative tip-sample distance $Z_{1st}$ regulated for a particular set-point {$\Delta f_{1}$} and acquiring a second pass along the same scan line with the feedback open and at a closer tip-sample distance $Z_{2nd}$ = $Z_{1st}$ - $Z_{off}$ ($Z_{off}$ is in the order of 200 to 400~pm). 

Figures~\ref{Fig2}c and~\ref{Fig2}d show such AFM images of the DCPDMbpy molecules on NiO(001) without and with Fe atoms, respectively. The two AFM images acquired during the first scan suggest that both molecules are lying flat on the surface ($\sim$ 0.2 nm in height) but the lack of resolution does not allow us to unambiguously assign a conformation.  In the second pass, the molecules are better resolved (Figures~\ref{Fig2}e~and~f) and a clear distinction between the $transoid$ and $cisoid$ conformations of both molecules is observed as illustrated in Figures~\ref{Fig2}g~and~h. Upon adsorption and without Fe atoms, the DCPDMbpy molecules are adsorbed in the prochiral $transoid$ conformation whereas successive deposition of Fe and molecules results in the formation of the Fe-DCPDM(bpy) units possessing the $cisoid$ conformation within the $bpy$ units. Although the adoption of the $cisoid$-conformation is due to the coordination to a metal centre, we cannot clearly confirm its presence by AFM imaging. Metal atoms are generally difficult to observe by AFM in metal-ligand complexes at surfaces~\cite{Kaw16,Koc16,Paw17a,Zin17,Kru18}.


\subsection{Structure Resolution by Low Temperature AFM With CO-Terminated Tips.} 
\begin{figure}[t]
 \centering
 \includegraphics[width=16cm]{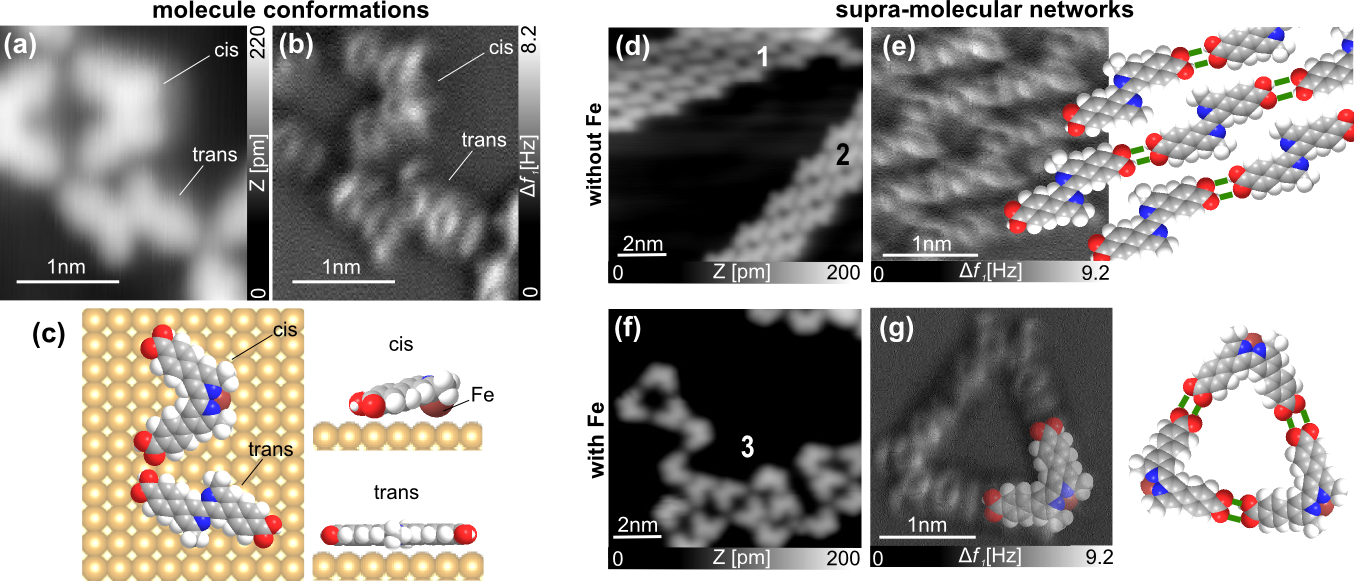}
 \caption{High-Resolution Imaging of The $Transoid$- And $Cisoid$- Conformations of DCPDMbyp And Fe-DCPDMbpy With CO-Terminated Tips. (a) STM image of the molecules in $transoid$- and $cisoid$- conformation. (b) Corresponding AFM image of the two same molecules with intra-molecular resolution. (c) Structural models of  DCPDMbyp in $transoid$- and  Fe-DCPDMbpy in $cisoid$-conformations. (d) Self-assemblies of $transoid$-DCPDMbpy on Au(111) leading to two enantiopure domains denoted $1$ and $2$, respectively. (e) Tentative structural model of H-bonded enantiopure molecular domains $1$ superimposed to an AFM image of the assembly with a CO-terminated tip. (f) STM image of the molecular network obtained by adding Fe atoms on Au(111). The pro-chirality of the molecule domain is lost due to the metal-complex formation.  (g) Tentative structural model an AFM image of the H-bonded Fe-DCPDMbpy molecules in their $cisoid$-conformation on Au(111). (Scan parameters: STM images: I$_t$ = 1 pA, V = -0.15 V and AFM images: $A$ = 50 pm, V = 0 V).}
   \label{Fig3}
 \end{figure}

To further improve the resolution, we measured the DCPCMbpy molecules on Au(111) at 4.7~K using AFM with a CO-terminated tip \cite{Gro09}. Compared to the NiO(001) samples, Au(111) surfaces were prepared with similar molecule and Fe atom coverages (see Methods). Figures~\ref{Fig3}a and \ref{Fig3}b show STM topographies of both DCPDMbpy conformations and the corresponding constant-height AFM image acquired with a CO-terminated tip at 4.7 K. The $transoid$- and $cisoid$-conformations are unambiguously observed and a clear distinction of the phenyl rings of the molecules as well as the methyl groups attached to the $bpy$ units again confirms that the molecules lie flat on the surface (see qualitative models in Figures~\ref{Fig3}c). For the $cisoid$-conformation, the Fe atom bound to the $bpy$ moieties is again not visible in the image. This observation, in addition to the fact that the methyl groups of the Fe-DCPDMbpy appear with a brighter contrast in comparison to $transoid$ -DCPDMbpy, suggest that the Fe atom is hidden under the molecule with the result that the latter undergoes a slight bending (Figures~\ref{Fig3}c).

The diffusion of the DCPDMbpy on Au(111) in comparison to what is observed on NiO(001) plays an important role. Indeed, in contrast to NiO, large DCPDMbpy self-assemblies can be formed at the gold surface at room temperature even at low coverage ($\leq 0.2$ ML) as shown in Figure~\ref{Fig3}d. In other words, the diffusion of each product of the reaction and consequently also the formation of supramolecular structures can be hindered or facilitated as function of the host substrate.
On Au(111), two enantiopure domains denoted $1$ and $2$ coexist as a direct consequence of the prochirality of the DCPDMbpy molecule. As shown in the AFM image and highlighted by green lines in the structural model of Figure~\ref{Fig3}e, the self-assembly process is governed by hydrogen bonding between carboxylic groups of adjacent enantiomers and forms extended close-packed molecular domains. The additional deposition of Fe atoms onto these chiral molecular domains on Au(111) leads to the formation of extended chain-like structures (Figure~\ref{Fig3}f and Figure~\ref{Fig4}d). By interacting with the lone pairs of the bipyridine unit, the Fe atom in the Fe-DCPDMbpy complex imposes the $cisoid$ conformation. This conformation is achiral on the surface and, thus, induces the loss of chirality of the molecular domains. 

In analogy to the self-assembly of $transoid$ molecules, the assembly of  $cisoid$ Fe-DCPDMbpy complex is driven by hydrogen bonding between their carboxylic groups (O-H...O) leading to the formation of chain-like assemblies (Figures~\ref{Fig3}f and \ref{Fig4}d) as well as trimers (Figure~\ref{Fig3}g). Note that the coordination of more than one DCPDMbpy ligand to Fe was never observed on the surface. We attribute this to the steric hindrance that would occur between the 6,6'-dimethyl groups of adjacent DCPDMbpy ligands in a Fe(DCPDMbpy)$_2$ species that was constrained to a planar conformation on a surface.

\subsection{XPS Study of The Complex on Au(111).}
\begin{figure}[t!]
 \centering
 \includegraphics[width=10cm]{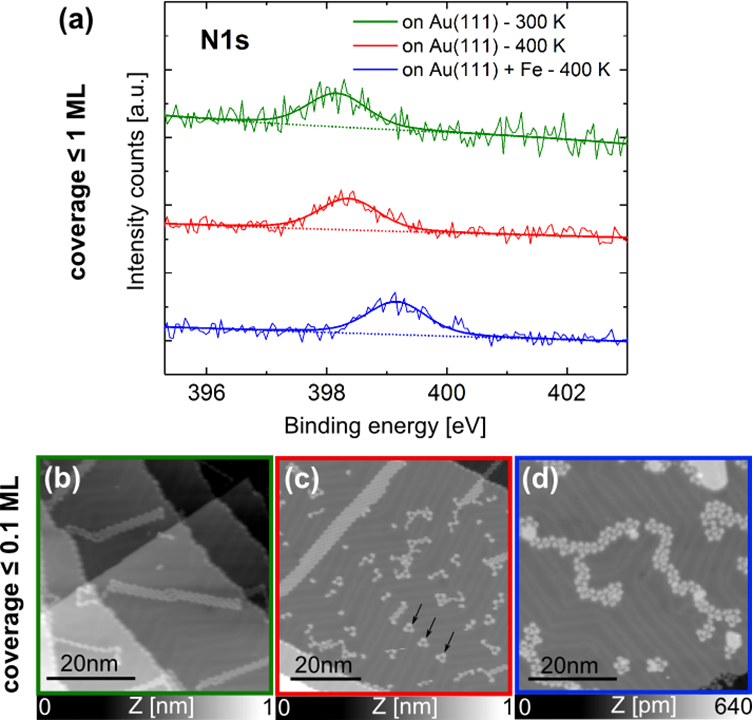}
 \caption{N1s Core Level Spectra. (a) XPS of one monolayer (ML) coverage of DCPDMbpy on Au(111) adsorbed at RT (green), after annealing at 400~K (red) and after annealing at 400~K and Fe deposition (blue). The shown spectra are normalized and shifted vertically for comparison. (b-d) Corresponding STM images of the surfaces after these preparations for coverage $\leq 1$~ML (scan parameters: I$_t$ = 1 pA, V = -0.15 V).} 
    \label{Fig4}
 \end{figure}

To investigate the role of Fe adatoms in the assembly process, we further investigated by XPS the N1s binding energies (BE) of the DCPDMbpy molecules on Au(111) to reveal the chemical environment of the $bpy$ moieties. The samples were prepared at RT with a coverage $\leq 1$~ML in the preparation chamber of the LT microscope and then transferred using a UHV-vacuum suitcase to the XPS chamber for analysis (see Methods). For this specific coverage, the N1s BE is at 398.2~eV  (green curve in Figure~\ref{Fig4}a) which corresponds to supramolecular networks of the $transoid$-molecules (Figure~\ref{Fig4}b). Upon complex formation obtained by additionally depositing Fe adatoms (blue curve in Figure~\ref{Fig4}a), the N1s BE significantly shifts by 1~eV to higher values (BE = 399.2~eV) supporting the expected  Fe complex formation and with this also the switch to $cisoid$-conformation. In that case, the lone-pair of the N atoms of the $bpy$ preferentially interact with an Fe adatom~\cite{Shc13,Mei17} inducing a new chemical environment  for the nitrogens (N...Fe...N). After annealing of the surface covered with DCPDM(bpy) molecules at 400~K (without Fe deposition), the N1s BE shifts slightly by 0.2~eV. According to STM image (Figure~\ref{Fig4}c), the shift originates from a fraction of DCPDMbpy molecules that have formed a complex with specific sites of the gold surface such as step edges, defects and elbows of the reconstruction  ( black arrows in Figure~\ref{Fig4}c)~\cite{Paw17a}. Although two peaks are expected here, the first arising from the transoid-DCPDMbpy molecules and the second from the Au-DCPDMbpy coordination complexes (N...Au...N), the small amount of molecules (<1ML), e.g. low signal-to-noise ratio, does not allow a proper deconvolution of the N1s peaks and hence  only one peak is observed. Moreover, the complex formation  reaction only triggered by temperature without additional metal atoms is less efficient since restricted to specific locations of the Au(111) surface.\\ 

Table~\ref{Tab1} summarizes the DCPDMbpy conformation ratio as a function of the sample preparations at a coverage $\leq 0.1$~ML adsorbed on Au(111). This ratio could be determined through analysis of a set of high resolution STM images. Upon deposition at RT, almost all the DCPCMbpy molecules adsorbed on the Au(111) surface are in $transoid$-conformation and the ratio $transoid$:$cisoid$ is measured to be $\sim$ 91:9. Upon Fe deposition and surface annealing, this ratio changes to 3:97 demonstrating a high specificity for the complex formation. When only triggered by temperature, the ratio becomes $\sim$ 52:48, but might be increased for higher molecule coverages since it will allow the saturation of the Au(111) reactive sites. It is also worth mentioning that metallic surfaces are known for their catalytic character in contrast to other surfaces. In a previous investigation, the study of the temperature effect on the DCPDMbpy assembly on NiO(001)~\cite{Fre17} demonstrated that molecules are not affected by annealing and remain in $transoid$-conformation up to 493 K when they tend to desorb. 
In principle, the process is thus independent of the underlying surface as soon as metal adatoms are present as demonstrated on both Au(111) and NiO(001) surfaces. As shown in our work, the diffusion and local reactive sites of the surfaces, however, influences the complex reaction and other metals might also form complexes. Finally, we emphasize the irreversible character of this prochiral to achiral transition in single molecules as well as in supramolecular networks  since the opposite change, from $cisoid$ to $transoid$, could not be experimentally achieved. Our results thus show the formation and suppression of surface-induced prochirality from the single molecule scale to the supramolecular network level.

\begin{table}
  \caption{Effect of Fe And Annealing on The DCPDMbpy Conformation For Less Than 0.1 Monolayer Coverage.}
  \label{Tab1}
\begin{tabular}{c c c c}
\hline
&  without Fe &  without Fe & with Fe \\
&  without annealing &  with annealing & with annealing\\
\hline
$transoid$ & 91\% & 52\% & 3\%\\
$cisoid$ & 9\% & 48\% & 97\%\\
\hline
\end{tabular}
\end{table}
\indent
 
\section{Conclusion}
In summary, the 4,4$'$-di(4-carboxyphenyl)-6,6$'$-dimethyl-2,2$'$-bipyridine molecule (DCPDMbpy) adsorbs in a $transoid$ geometry on both NiO(001) and  Au(111) as single molecules and enantiopure domains, respectively. When adsorbed on NiO(001) partially covered with Fe adatoms, the molecule shows a $cisoid$ conformation demonstrating the formation of a metal-coordination complex (Fe-DCPDMbpy). On Au(111), we showed that the molecules undergo the interconversion from $transoid$ to $cisoid$ upon Fe adatom deposition on previously formed enantiopure  DCPDMbpy assemblies. Using  AFM imaging and XPS measurements, we demonstrated that the process is triggered by coordination complex formation between Fe atoms and the $bpy$ moieties of the molecule. Interestingly, the new  Fe-DCPDMbpy supra-molecular networks on gold are achiral, which demonstrates the suppression of a surface-induced chirality in thin supramolecular networks via metal complex formation.

\section{Methods}

\subsection{Molecule Synthesis}
DCPDMbpy was synthesized by Dr. Davood Zare (University of Basel) following the reported procedure \cite{Her09}..

\subsection{Sample Preparation}
The NiO(001) crystals used in this study, purchased from SurfaceNet, consist of a rectangular rod with dimensions $2\times2\times7$ mm$^{3}$ and a long axis in the [001] direction. The NiO(001) surface was prepared through {\it in-situ} cleavage ( UHV, $p<1\times10^{-10}$~mbar) with prior and subsequent annealing at about 800 K resulting in an atomically clean surface. An Au(111) single crystal, purchased from Mateck GmbH, was cleaned by several sputtering and annealing cycles in  UHV conditions. DCPDMbpy molecules were thermally evaporated from a Knudsen cell heated up to 528~K on the surfaces kept at RT. The molecule rate was checked {\it in-situ} using a quartz micro-balance. Fe adatom depositions were conducted using an e-beam evaporator. To promote complex formation, the sample was then annealed to 420~K during molecule and atom evaporation. For NiO(001), because of the low diffusion rate, we first sublimated the Fe atoms and then  DCPDMbpy molecules. For Au(111), the steps of the procedure were inverted: molecules were deposited first and Fe atoms  afterwards. A vacuum suitcase from Ferrovac GmbH was employed to transfer samples from the UHV LT AFM/STM setup to the XPS chamber.

\subsection{AFM Imaging at Room Temperature}
AFM measurements on NiO were conducted with a home built atomic force microscope (AFM) in UHV operated at RT. All AFM images were recorded in the non-contact mode (nc-AFM), using silicon cantilever (Nanosensors PPP-NCR stiffness $k = 20-30$~N/m, resonance frequency $f_{1}$ around 165~kHz and $Q_{1}$ factor around 30000  with compensated contact potential difference (CPD). 
 
\subsection{STM/AFM Imaging at Low Temperature}
STM/AFM experiments were carried out at 4.7~K with an Omicron GmbH low-temperature STM/AFM operated with a Nanonis RC5 electronics. We used commercial tuning fork sensors in the qPlus configuration ($f_1$~=~26~kHz, $Q$~= 10000-25000, nominal spring constant k = 1800~N.m$^{-1}$). The constant-height AFM images were acquired with CO-terminated tips. All voltages refer to the sample bias with respect to the tip.

\subsection{XPS Measurements}
The samples were transferred {\it in-situ} using a vacuum suitcase to the XPS chamber directly after molecules and Fe atom deposition. The pressure in the XPS chamber was always $\leq$~10$^{-10}$ mbar and measurements were performed using a VG ESCALAB 210 system equipped with a mono-chromatic Al K$_{\alpha}$ radiation source. A pass energy of 20~eV was used for all narrow scan measurements and 100~eV pass energy for survey scans. Normal electron escape angle and a step size of 0.05~eV were used. The energy positions of the spectra were calibrated with reference to the 4f~7/2 level of a clean gold sample at 84.0~eV binding energy. XPS fitting was performed with Unifit 2016 Software \cite{Hesse99}.

\section{Author Contributions}
S.F., A.H., C.E.H. and T.G. conceived the experiment. S.F. measured the RT-AFM on NiO, R.P. measured with the LT-STM/AFM on Au(111). L.M. and R.S. performed the XPS measurements. S.F. wrote the manuscript with the help of R.P. All co-authors contributed to project concepts, discussion and read and commented on the manuscript.

\begin{acknowledgement}

This work was supported by the Swiss National Science Foundation (SNF) CR22I2-156236, the Swiss Nanoscience Institute (SNI) and the University of Basel.

\end{acknowledgement}






\end{document}